\documentclass[10pt,letterpaper,twocolumn]{article}

\usepackage[T1]{fontenc}
\usepackage[margin=0.75in]{geometry}
\usepackage{mathptmx}
\usepackage[scaled=0.95]{helvet}
\usepackage{amsmath}
\usepackage{graphicx}
\graphicspath{{./}}
\usepackage{booktabs,array}
\usepackage{microtype}
\usepackage{needspace}
\usepackage{placeins}
\usepackage{titlesec}
\usepackage[super]{cite}
\usepackage[font=small,labelfont=bf,labelsep=period,
  justification=justified,singlelinecheck=false]{caption}
\usepackage[hidelinks]{hyperref}

\titleformat{\section}{\sffamily\bfseries\normalsize}{}{0pt}{\MakeUppercase}
\titleformat{\subsection}[runin]{\bfseries}{}{0pt}{}[.]
\titlespacing*{\section}{0pt}{10pt plus 2pt minus 1pt}{5pt}
\titlespacing*{\subsection}{\parindent}{6pt plus 1pt}{0.5em}
\title{Explainable Molecular Structure Inference from GC--MS with
Diffusion Models and LLM Reranking}
\author{Changlin Liu, Tianyu Yi, Chengchun Liu, Boxuan Zhao, and Fanyang
Mo\textsuperscript{*}}
\date{}

\makeatletter
\renewcommand{\@maketitle}{%
  \noindent{\raggedright\sffamily\bfseries\hyphenpenalty=10000
    \fontsize{17}{20}\selectfont\@title\par}%
  \vspace{8pt}\noindent{\normalfont\normalsize\@author\par}
  \vspace{10pt}%
}
\makeatother

\renewenvironment{abstract}{%
  \noindent\begin{minipage}{\textwidth}\normalfont\normalsize
  \textbf{ABSTRACT:}\enspace\ignorespaces
}{\par\end{minipage}\par}

\begin{document}

\twocolumn[{%
\begin{@twocolumnfalse}
\maketitle
\begin{abstract}
GC--EI--MS is an important technique for analyzing volatile and semivolatile
compounds in complex samples. However, conventional methods rely heavily on
reference spectral library matching, limiting their ability to identify
compounds absent from these libraries and to infer complete molecular
structures directly from fragmentation information. Here, we present DiffGCMS,
a spectrum-conditioned discrete graph diffusion model for de novo structure
elucidation from GC--EI--MS, and further develop a framework that integrates
DiffGCMS with second-stage reasoning by a large language model (LLM). In the
first stage, DiffGCMS generates candidate molecular structures from input
spectra; in the second stage, the LLM uses mass spectral information to
validate, repair, and rerank the candidates and provides interpretable analysis
of fragment-ion peaks. This framework can generate plausible molecular
structures for compounds absent from reference spectral libraries and provide
traceable evidence supporting its decisions. On a test set comprising 13,696
spectra from NIST 20, the generative model achieved Acc@1 and Acc@10 of 6.01\%
and 15.76\%, respectively. On the test subset containing molecules with no more
than 10 heavy atoms, LLM-assisted molecular graph repair and reranking increased
Acc@1 from 21.28\% to 21.95\%, Acc@10 from 46.91\% to 47.99\%, and candidate
validity from 91.04\% to 100\%. These results demonstrate that spectrum-aware
postprocessing can correct errors produced by the generative model while
providing auditable and traceable explanations for the final ranking.
\end{abstract}
\vspace{10pt}
\end{@twocolumnfalse}
}]

\begin{figure*}[!t]
  \centering
  \includegraphics[width=\textwidth,trim={145.20bp 191.54bp 27.48bp 208.81bp},clip]{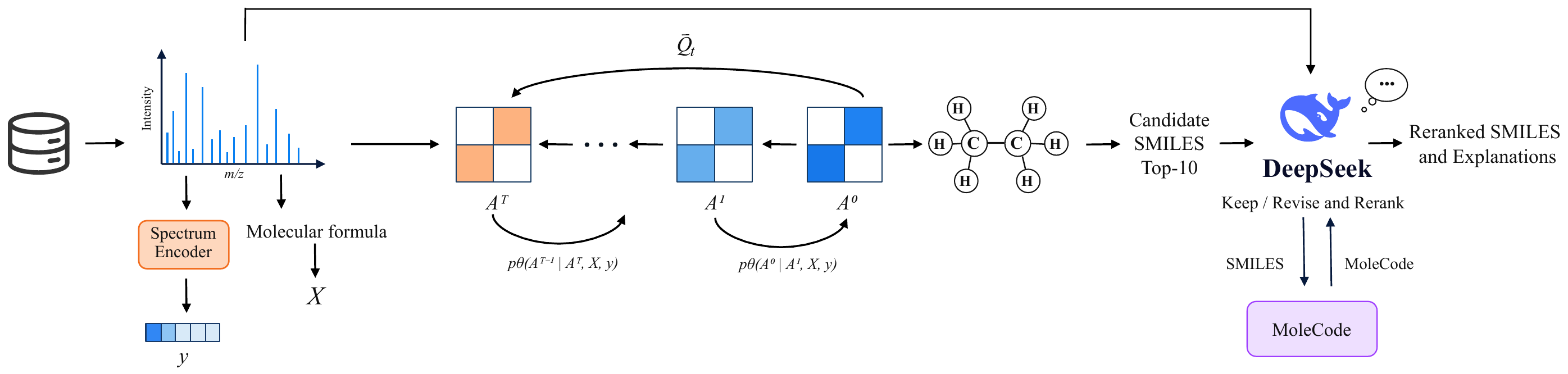}
  \caption{\textbf{Overall two-stage workflow for molecular structure inference from
  GC--EI--MS.} The query spectrum is encoded as the conditioning vector $y$, and
  the molecular formula determines the heavy-atom set $X$. DiffGCMS progressively denoises
  the bond-adjacency state from $A^{T}$ through $A^{t}$ to the predicted clean
  molecular graph $A^{0}$. Ten sampling runs are performed for each spectrum to
  generate 10 candidate SMILES. In the second stage, DeepSeek uses bidirectional SMILES--MoleCode conversion to
  inspect, retain or revise, and rerank the candidates, returning ranked SMILES
  together with mass-spectral explanations.}
  \label{fig:workflow}
\end{figure*}

Gas chromatography--mass spectrometry (GC--MS) is a widely used analytical
platform for characterizing volatile and semivolatile compounds in
environmental analysis, food safety, forensic science, metabolomics, and
industrial quality control. In a typical GC--MS experiment, compounds are first
separated by gas chromatography and subsequently ionized; the resulting ions
are detected according to their mass-to-charge ratios ($m/z$) and relative
abundances. Electron ionization (EI), typically operated at a standard
ionization energy of 70~eV, produces highly reproducible, fragment-rich mass
spectra. These characteristics have facilitated the construction of large
reference spectral libraries and made gas chromatography--electron ionization
mass spectrometry (GC--EI--MS) particularly suitable for routine compound
identification across different laboratories and instruments
\cite{nist20,stein2012,stein1994}.

Conventional compound identification by GC--EI--MS relies primarily on
spectral similarity algorithms that compare experimentally measured spectra
with standard reference spectra\cite{stein1994,stein1999,massbank}. These methods perform well when the compound
of interest is represented in the spectral library. However, identification
performance decreases substantially when the analyte is absent from the
reference library. More broadly, in silico fragmentation and spectrum prediction can extend
mass-spectral identification beyond experimentally measured spectral libraries.
In database-search implementations, however, identification remains restricted
to the supplied candidate structures\cite{metfrag,cfmei,krettler2021}.

De novo molecular generation offers a promising alternative by directly
inferring molecular structures from mass spectra without requiring the target
structure to be present in a predefined database. Most existing studies in
this area have focused on liquid chromatography--tandem mass spectrometry
(LC--MS/MS). Models such as Spec2Mol\cite{spec2mol},
MassGenie\cite{massgenie}, MSNovelist\cite{msnovelist},
MS2Mol\cite{ms2mol}, and Mass2SMILES\cite{mass2smiles} use neural
sequence-generation or learned molecular-representation approaches to infer
candidate structures from tandem mass spectra and, depending on the method,
molecular formulas or predicted fingerprints.
Cycle-MS integrates spectrum-to-molecule prediction with forward spectrum
reconstruction through cycle-consistency training for de novo structure
elucidation from tandem mass spectra\cite{cyclems}.
Graph-based approaches such as DiffMS\cite{diffms} and
FlowMS\cite{flowms} generate candidate molecular structures through iterative
diffusion or flow-matching processes under molecular-formula constraints. LC--MS/MS can also exploit
precursor-ion information and multiple fragmentation conditions to facilitate
the inference of molecular masses or molecular formulas. Consequently, de novo
structure elucidation from LC--MS/MS has developed into a relatively active and
mature research area.

In contrast, direct de novo molecular generation from GC--EI--MS remains
comparatively underexplored. Previous machine-learning studies have focused
mainly on forward EI spectrum prediction or database-dependent retrieval and
identification. NEIMS\cite{neims} and RASSP\cite{rassp} predict EI mass spectra
from molecular structures, whereas DeepEI\cite{deepei} maps experimental
spectra to molecular fingerprints for candidate ranking. Although these
approaches improve the coverage
or efficiency of spectral-library searching, they do not fully eliminate
dependence on candidate databases. SpecTUS\cite{spectus} represents an important advance: it
employs a large encoder--decoder Transformer to translate low-resolution EI
mass spectra directly into SMILES representations. The model is first
pretrained on millions of synthetic spectra and then fine-tuned on experimental
NIST spectra, demonstrating the feasibility of generating molecular structures
directly from EI--MS data. Nevertheless, autoregressive SMILES generation may
produce chemically implausible structures or structures that lack sufficient
support from spectral evidence. Sequence-generation scores do not by themselves provide an explicit
peak-by-peak account of how a proposed structure explains the observed EI
spectrum.

Here, we present DiffGCMS, a spectrum-conditioned discrete graph diffusion
framework for de novo molecular structure elucidation from GC--EI--MS data. The
model represents atoms and chemical bonds as categorical variables and
iteratively denoises molecular graphs under molecular-formula constraints,
enabling the generation of chemically constrained and structurally diverse
molecular candidates. To bridge the gap between generating valid molecules and
selecting the true structure best supported by experimental evidence, we
further develop a second-stage reasoning framework based on a large language
model (LLM). For each query spectrum, DiffGCMS first generates a set of
candidate molecular structures. The LLM subsequently evaluates these
candidates against molecular-ion and fragment-ion evidence, performs
chemistry-rule-guided molecular graph correction when necessary, reranks the
candidate set, and provides peak-level explanations for its decisions. By
decoupling molecular generation from evidence-based validation, this framework
combines the generative capability of discrete diffusion models with the
chemical reasoning and interpretability of LLMs, providing a traceable strategy
for identifying compounds beyond existing GC--EI--MS spectral libraries. The
overall workflow is illustrated in Figure~1.

\section{Methods}

\begin{figure*}[!t]
  \centering
  \includegraphics[width=\textwidth,trim={126.30bp 158.41bp 256.71bp 18.57bp},clip]{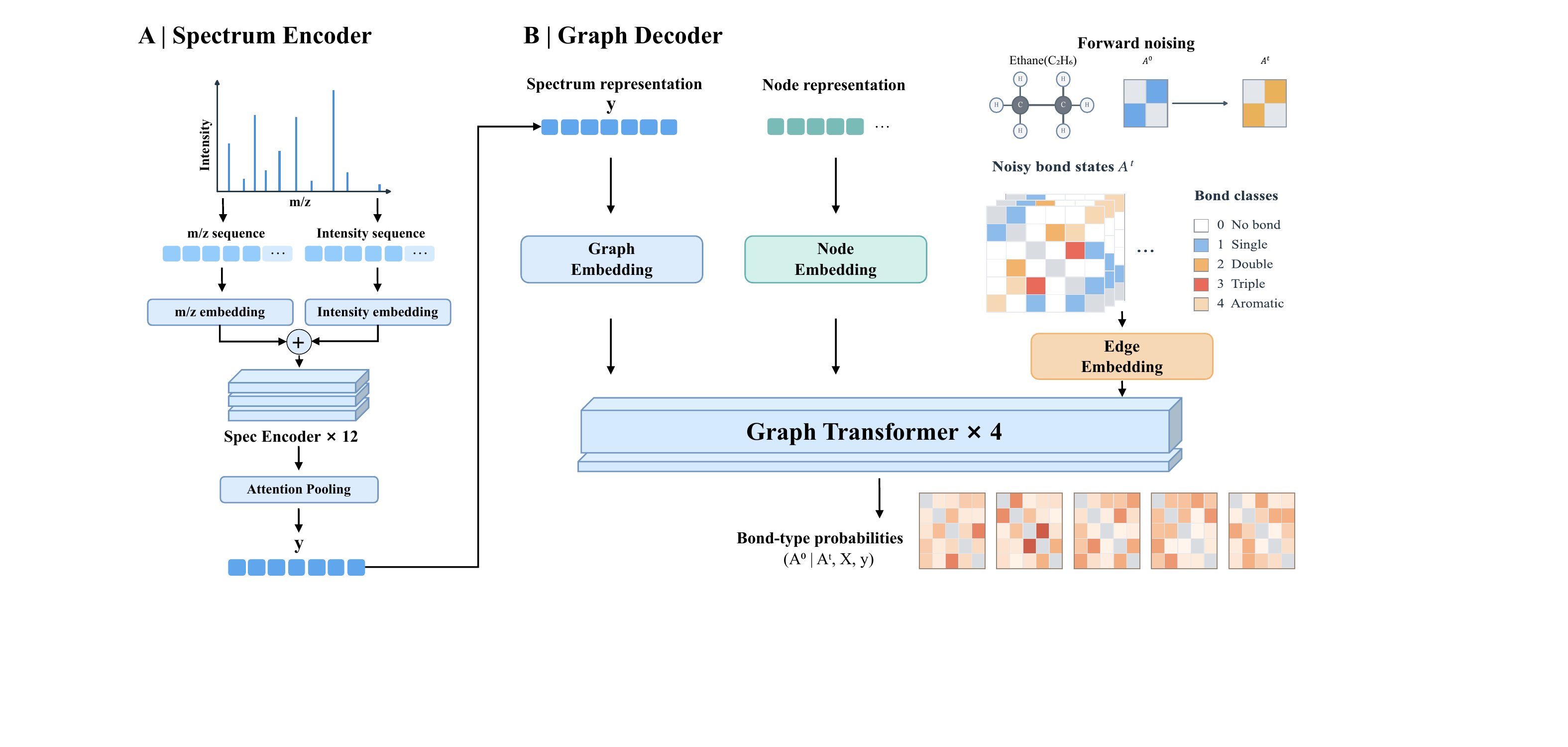}
  \caption{Architecture of DiffGCMS. (A) The spectrum encoder embeds the
  mass-to-charge ratio ($m/z$) and intensity sequences and adds the two embeddings
  to obtain peak representations. A 12-layer spectrum encoder and a pooling
  module then produce the global spectrum conditioning vector $y$. (B) The graph
  decoder combines the spectrum condition, the heavy-atom types, counts, and
  structural features determined by the molecular formula, and the noisy bond
  state $A^{t}$. Four GraphTransformer layers predict probabilities for five bond
  classes (no bond, single, double, triple, and aromatic). During sampling, the
  predicted distribution is used to iteratively update the bond state.}
  \label{fig:model}
\end{figure*}

\subsection{Dataset Collection and Preprocessing}
This study used gas chromatography--electron ionization mass spectrometry
(GC--EI--MS) data from the NIST 20 Mass Spectral Library\cite{nist20}. To ensure data quality
and compatibility with model input requirements, the raw data were subjected
to standardized filtering. A sample was excluded if it met any of the following
criteria: (1) missing retention index (RI); (2) molecular weight greater than
600~Da; (3) presence of elements other than C, H, N, O, S, P, F, Cl, Br, I, Si,
and B; (4) more than 300 peaks in the mass spectrum; (5) maximum mass-to-charge
ratio ($m/z$) greater than 500; or (6) a corresponding molecular structure that
could not be parsed by RDKit (version 2022.09.5)\cite{rdkit} or failed chemical validity checks. After
filtering, 273,935 valid GC--EI--MS spectrum--structure pairs were retained.

The molecular structures were then standardized using RDKit, and the SMILES
strings were converted to canonical SMILES for molecular identity assignment
and dataset partitioning\cite{smiles}. To prevent data leakage caused by spectra of the same
molecule appearing in both the training and test sets, we adopted a
molecule-disjoint splitting strategy. Samples were first grouped by canonical
SMILES and then divided into training, validation, and test sets at approximate
ratios of 90\%, 5\%, and 5\%, respectively.

\subsection{DiffGCMS: Spectrum-Conditioned Graph Diffusion}
DiffGCMS is a spectrum-conditioned discrete graph diffusion model for de novo
molecular structure inference from GC--EI--MS. It consists of a spectrum
encoder, a condition projection module, and a graph diffusion decoder. The
model takes a mass spectral peak sequence and a molecular formula as inputs.
The spectrum provides fragment-ion information, whereas the molecular formula
determines the number and types of nodes in the molecular graph. Rather than
jointly generating atoms and chemical bonds, DiffGCMS fixes the atom set
specified by the molecular formula and applies discrete diffusion only to the
interatomic connectivity, thereby constraining the generation space to
candidate structures consistent with the given formula.

The spectrum encoder is initialized from the pretrained SpecTUS encoder\cite{spectus}. For
each spectral peak, the integer-valued mass-to-charge ratio is converted into
an $m/z$ embedding, while the normalized intensity is logarithmically binned
and converted into an intensity embedding. The two embeddings are added to
form a peak-level input representation. The resulting peak sequence is then
processed by a 12-layer Transformer encoder with an embedding dimension of
1024, a feed-forward dimension of 4096, and 16 attention heads, enabling the
model to capture global relationships among fragment peaks. The encoder outputs
are aggregated by mask-aware mean pooling to obtain a fixed-length spectrum
representation. This representation is subsequently passed through a two-layer
condition projection module with dimensions of 1024, 4096, and 1024. A SiLU
activation function is applied between the two linear layers, producing a
1024-dimensional global spectrum-conditioning vector.

A molecular structure is represented as an undirected graph in which nodes
correspond to heavy atoms and edges represent chemical bonds. Node types are
determined by the molecular formula, whereas edges are classified into five
categories: no bond, single bond, double bond, triple bond, and aromatic bond.
The forward diffusion process is applied exclusively to the edge variables.
During each training iteration, a diffusion time step is uniformly sampled
from 1 to 500, and the corresponding noisy edge structure is sampled from the
ground-truth graph using the cumulative transition matrix. The diffusion
process comprises 500 steps and follows a cosine noise schedule, following
established discrete diffusion and graph-denoising formulations
\cite{d3pm,digress,nichol2021}. The categorical
transition matrix is constructed from the empirical distribution of edge types
in the training set. Because no noise is applied to the nodes, the atom types
and spectrum-conditioning vector remain unchanged throughout the diffusion
process.

The graph diffusion decoder consists of four GraphTransformer layers
\cite{graphtransformer,digress}. Its
inputs include the fixed node types, the noisy edges at the current diffusion
step, the spectrum-conditioning vector, the normalized diffusion time, and
additional structural and chemical features. These auxiliary features include
node valence, formal charge, local ring information, connected-component
information, and graph Laplacian spectral features, which enhance the model's
awareness of molecular topology and chemical constraints. The node, edge, and
global features are first mapped into their respective hidden spaces using
multilayer perceptrons and are then jointly updated by the GraphTransformer.

Each GraphTransformer layer comprises a joint node--edge attention module,
residual connections, layer normalization, and feed-forward networks. Node
attention first models interactions between atoms, after which the current edge
features multiplicatively and additively modulate the attention
representations, allowing the bond states to directly influence information
exchange between nodes. The spectrum condition is injected into the node and
edge representations through feature-wise linear modulation, ensuring that
every denoising step is guided by the input mass spectrum. Aggregated node and
edge representations are also used to update the global features, enabling
information exchange among the spectrum, atoms, and chemical bonds. The output
layer predicts the probability of each edge type for every pair of atoms in the
clean molecular graph. The predicted edge matrix is symmetrized, and its
diagonal elements are masked to ensure that the generated structure is an
undirected graph without self-loops. The overall model architecture is shown in
Figure~\ref{fig:model}.

During training, the ground-truth edge types in the original molecular graph
are used as supervision, and the edge-classification cross-entropy loss is
computed only over valid node pairs. Because the adjacency matrix is symmetric,
the loss is evaluated only over its upper-triangular entries to avoid counting
each undirected edge twice. Neither the node types nor the global conditioning
features contribute directly to the training loss. The model is optimized using
AdamW together with a OneCycle learning-rate schedule\cite{adamw,onecycle}.

During sampling, the node types and spectrum condition remain fixed, while the
initial edge structure is sampled from the empirical marginal distribution of
edge types in the training set. Starting from diffusion step 500, the model
progressively predicts the edge-type distribution of the clean molecular graph
and samples the edge structure at the preceding step using the discrete
diffusion posterior. This procedure continues until the complete
reverse-denoising process is finished. The resulting discrete graph is
converted into a candidate molecular structure using RDKit. For each input
spectrum, the sampling procedure is repeated to generate 10 candidate
structures for subsequent validation and reranking by the large language
model.

\begin{figure*}[!t]
  \centering
  \includegraphics[width=\textwidth,trim={148.98bp 52.58bp 76.22bp 56.36bp},clip]{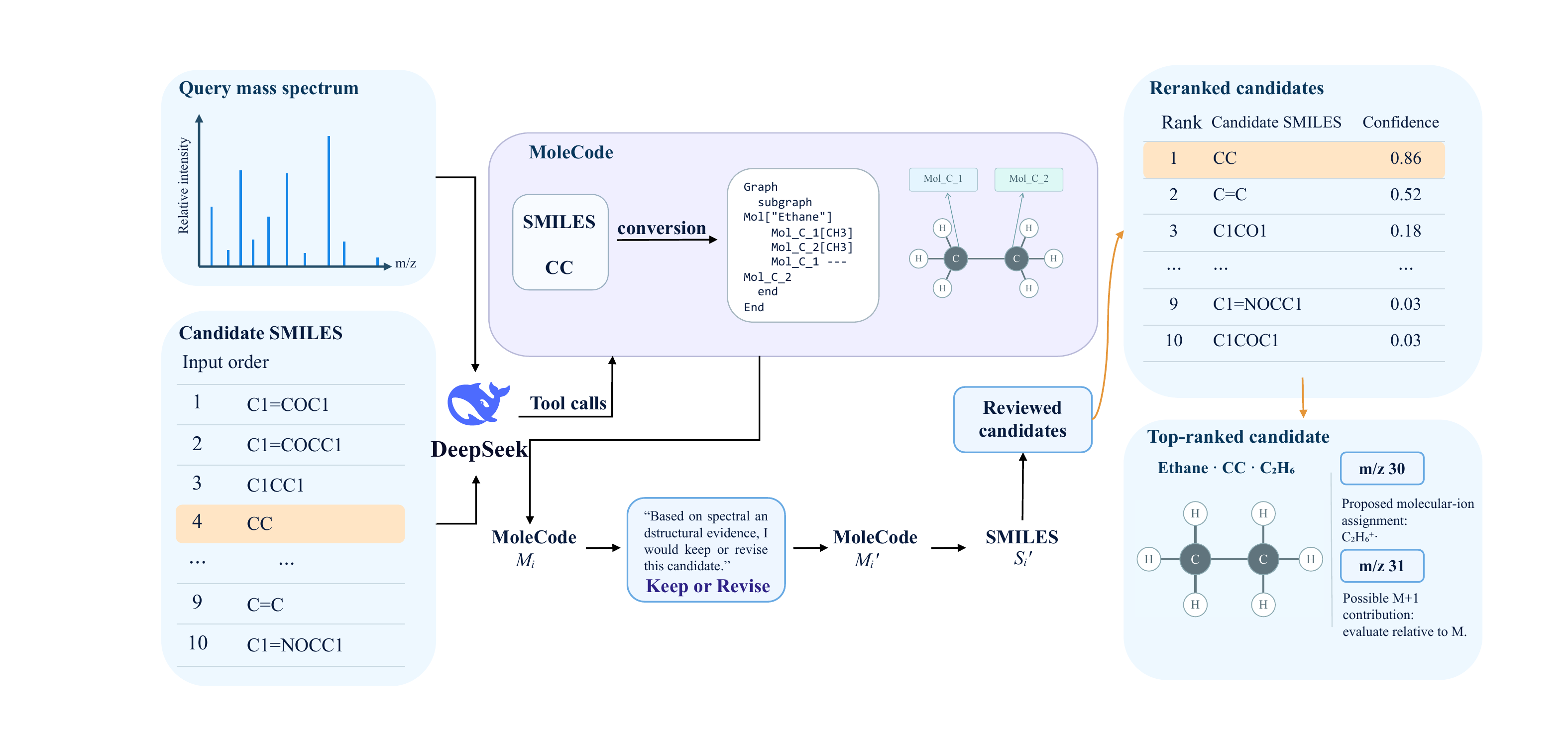}
  \caption{MoleCode-enabled LLM graph repair and spectrum-aware reranking. The
  query EI spectrum and 10 DiffGCMS candidate SMILES are supplied to DeepSeek.
  For each candidate, the model invokes deterministic SMILES-to-MoleCode
  conversion, inspects the explicit atom--bond graph, and retains or locally
  revises the structure using spectral and chemical evidence. Reviewed
  MoleCode graphs are converted back to SMILES, validated, assigned confidence
  scores, and reranked. Peak-level interpretation is then reported for the
  highest-confidence candidate.}
  \label{fig:llm}
\end{figure*}

\Needspace{6\baselineskip}
\subsection{Graph-Explicit Candidate Representation with MoleCode}
SMILES provides a compact and widely supported serialization for molecular
storage and exchange, but it presents a molecular graph as a one-dimensional
sequence\cite{smiles}. Connectivity is therefore implicit: branches are encoded through
parentheses, ring closures through paired indices, and bond relationships
through syntax whose meaning depends on position. An LLM must reconstruct the
underlying adjacency before it can compare structures or evaluate a proposed
chemical edit. This extra step is particularly problematic for the present
workflow, where candidate validation requires atom-resolved comparisons,
localized bond-order corrections, and a traceable record of which structural
feature supports a decision. Reconstruction errors can cause the model to
misidentify the target atom or bond even when the input SMILES is chemically
valid. The MoleCode study found that this representational bottleneck becomes
more pronounced for unfamiliar molecules, larger structures, and
topology-sensitive operations\cite{molecode}.

MoleCode addresses this limitation with a training-free, graph-explicit
molecular language built from Subgraph, Node, and Edge primitives. Atoms and
fragments are represented as typed entities with persistent identifiers, while
bonds are stated as explicit relations between nodes. Molecular topology is
therefore directly readable and editable in the language context rather than
being recovered from sequential syntax. MoleCode also supports deterministic
bidirectional conversion with standard formats such as SMILES and MOL, allowing
the graph representation to be introduced without changing the underlying
chemical structure. In our workflow, all candidate strings generated by DiffGCMS are retained and supplied directly to the large language model, including candidates that fail RDKit chemical validity checks. The original candidate strings remain available to the LLM for inspection and repair. During reasoning, the model uses a SMILES-to-MoleCode conversion tool to obtain explicit molecular graphs for convertible candidates. This graph-explicit interface provides a natural bridge between the graph diffusion generator and the subsequent evidence-aware reasoning stage.

\subsection{LLM Graph Repair and Spectrum-Aware Reranking}
The second stage receives the mass spectral peak list and 10 candidate SMILES
generated by DiffGCMS. DeepSeek-V4-Flash is used as the reasoning model in this
stage.

The LLM first reads all 10 candidate strings, including chemically invalid
candidates, and uses a SMILES-to-MoleCode conversion tool during reasoning to
support graph-level inspection. The original strings are retained as inputs
for structural repair regardless of their initial RDKit validation outcome.
MoleCode records atom-level and
bond-level information, including atom identity, hydrogen state, atomic
connectivity, and bond order, thereby making the molecular topology and chemical
properties implicit in the SMILES syntax explicitly accessible for
inspection\cite{molecode}. This tool-augmented organization is consistent with
prior chemistry-agent work showing the value of coupling LLM reasoning to
deterministic chemical tools\cite{chemcrow}.

Using the candidate strings and available MoleCode representations, the LLM jointly evaluates the
chemical plausibility of each candidate and its consistency with the input
spectrum by considering the nominal molecular mass, base peak, abundant
diagnostic ions, and chemically plausible neutral losses. Based on the
substructural information and bond-cleavage patterns reflected by the fragment
ions, the model identifies local structural features that are inconsistent with
the spectral evidence and further examines the corresponding regions for
valence violations, incorrect aromaticity assignments, implausible bond-order
patterns, or errors in ring connectivity. For recoverable molecular graph
errors, the LLM
proposes structural repairs while preserving the heavy-atom composition of the
original candidate.

The inspected or repaired candidates are returned as SMILES and subjected to
RDKit chemical validity checks, together with checks of molecular connectivity,
heavy-atom composition, and graph-to-SMILES round-trip consistency where
applicable. For each validated candidate, the LLM integrates molecular-mass consistency,
diagnostic-ion matches, plausible bond-cleavage patterns, neutral-loss
pathways, and supporting or contradictory evidence to assign a confidence score
and rerank the candidate structures accordingly. After reranking, the LLM
performs a detailed peak-level interpretability analysis only for the
highest-confidence Top-1 candidate, including the likely origins of key
fragment ions, the corresponding bond-cleavage pathways, and plausible neutral
losses. The complete workflow of tool invocation, molecular graph repair,
chemical validation, candidate reranking, and Top-1 peak-level interpretation
is illustrated in Figure~\ref{fig:llm}.

\subsection{Evaluation Metrics}
Let $N$ denote the number of query spectra, $G_i^*$ the standardized reference
structure for spectrum $i$, and $\widehat{G}_{ij}$ the candidate ranked at
position $j$. The function $c(\cdot)$ returns the standardized canonical-SMILES
identity, $\boldsymbol{f}(\cdot)$ returns a binary Morgan fingerprint, and
$\mathbf{1}[\cdot]$ is the indicator function\cite{ecfp}. Acc@$k$ was the fraction of
spectra for which the reference structure occurred within the first $k$
candidates:
\begin{equation}
\operatorname{Acc}@k = \frac{1}{N}\sum_{i=1}^{N}
\max_{1\leq j\leq k}\mathbf{1}\!\left[
c(\widehat{G}_{ij})=c(G_i^*)\right].
\label{eq:acc}
\end{equation}
Tanimoto@$k$ was the mean, over spectra, of the highest Morgan-fingerprint
Tanimoto similarity within the first $k$ candidates:
\begin{equation}
\operatorname{Tanimoto}@k = \frac{1}{N}\sum_{i=1}^{N}
\max_{1\leq j\leq k}T\!\left(
\boldsymbol{f}(G_i^*),\boldsymbol{f}(\widehat{G}_{ij})\right),
\label{eq:tanimoto-k}
\end{equation}
where the Tanimoto coefficient for two binary fingerprint vectors
$\boldsymbol{a}$ and $\boldsymbol{b}$ is
\begin{equation}
T(\boldsymbol{a},\boldsymbol{b})=
\frac{\boldsymbol{a}^{\mathsf T}\boldsymbol{b}}
{\lVert\boldsymbol{a}\rVert_1+\lVert\boldsymbol{b}\rVert_1-
\boldsymbol{a}^{\mathsf T}\boldsymbol{b}}.
\label{eq:tanimoto}
\end{equation}
For validity, let $M_i$ be the number of generated or postprocessed candidates
evaluated for spectrum $i$, and let $v(\widehat{G}_{ij})$ equal 1 when a
candidate passes all chemical checks and 0 otherwise. Validity was calculated
as
\begin{equation}
\operatorname{Validity}=
\frac{\displaystyle\sum_{i=1}^{N}\sum_{j=1}^{M_i}
v(\widehat{G}_{ij})}
{\displaystyle\sum_{i=1}^{N}M_i}.
\label{eq:validity}
\end{equation}
Results were reported for the full test set and for subsets containing no more
than 20 or 10 heavy atoms. The heavy-atom subsets quantify how molecular size
affects the combinatorial graph search problem and permit a focused evaluation
of the more expensive LLM stage.

\section{Results and Discussion}

\begin{table*}[!t]
  \caption{DiffGCMS performance on the full NIST 20 test set and the two
  heavy-atom subsets. Counts in parentheses are numbers of spectra for which
  the reference structure was exactly recovered.}
  \label{tab:generator}
  \centering
  \small
  \begin{tabular}{lccc}
    \toprule
    Metric & Full test set & Heavy atoms $\leq 10$ & Heavy atoms $\leq 20$ \\
    \midrule
    Number of spectra & 13,696 & 1,490 & 8,729 \\
    Number of generated structures & 136,960 & 14,900 & 87,290 \\
    Acc@1 & 6.01\% (823) & 21.28\% (317) & 8.56\% (747) \\
    Acc@10 & 15.76\% (2,158) & 46.91\% (699) & 22.37\% (1,953) \\
    Tanimoto@1 & 30.39\% & 39.94\% & 32.92\% \\
    Tanimoto@10 & 45.77\% & 63.37\% & 50.21\% \\
    Validity & 73.67\% & 91.04\% & 79.01\% \\
    \bottomrule
  \end{tabular}
\end{table*}

\subsection{DiffGCMS Candidate Generation}
Table~\ref{tab:generator} summarizes the molecular structure generation
performance of DiffGCMS on the test set. The reference structure was ranked
first for 823 spectra, corresponding to an Acc@1 of 6.01\%. For 2,158 spectra,
the reference structure was recovered within the set of 10 generated
candidates, yielding an Acc@10 of 15.76\%. The corresponding Tanimoto@1 and
Tanimoto@10 values were 30.39\% and 45.77\%, respectively. Compared with the
exact-match accuracies, the higher fingerprint similarities indicate that,
even when the reference molecule was not fully reconstructed, some generated
candidates retained substructural features similar to those of the reference,
such as local functional groups, ring systems, or molecular scaffolds. These
structurally related candidates may therefore remain informative for downstream
structure screening. The overall chemical validity of the generated structures
was 73.67\%, indicating that most candidates could be parsed as chemically valid
molecules, although some contained valence violations, inconsistent bond
orders, or implausible atomic connections.

Model performance was strongly dependent on molecular size. The subset of
molecules containing no more than 10 heavy atoms comprised 1,490 spectra and
14,900 generated structures. Within this subset, Acc@1 reached 21.28\%, while
Acc@10 increased to 46.91\%. The corresponding Tanimoto@1 and Tanimoto@10
values were 39.94\% and 63.37\%, respectively, and the chemical validity reached
91.04\%. Compared with the full test set, Acc@1 and Acc@10 increased by 15.27
and 31.15 percentage points, respectively, demonstrating that DiffGCMS more
effectively recovers the molecular connectivity of smaller compounds. In
particular, the Acc@10 value approaching 50\% indicates that multicandidate
sampling can effectively explore the chemical space surrounding the reference
structure for molecules with relatively few atoms. For the subset containing
molecules with no more than 20 heavy atoms, Acc@1 and Acc@10 were 8.56\% and
22.37\%, respectively. Tanimoto@1 and Tanimoto@10 were 32.92\% and 50.21\%,
respectively, and the chemical validity was 79.01\%.

Overall, Acc@1, Acc@10, Tanimoto similarity, and chemical validity all decreased
as larger molecules were included in the evaluation set. This trend can
primarily be attributed to the rapid expansion of the molecular graph search
space with increasing numbers of nodes. As molecular size increases, the number
of possible atom pairs grows approximately quadratically, while the number of
possible bond configurations and chemically plausible isomers increases even
more rapidly. Consequently, the reverse diffusion process must recover the
correct topology from a substantially larger set of possible structures.
Larger molecules also tend to contain more complex branching patterns, ring
systems, and combinations of bond orders, increasing the likelihood of local
valence violations and connectivity errors. The decrease in chemical validity
from 91.04\% for molecules with no more than 10 heavy atoms to 79.01\% for those
with no more than 20 heavy atoms, and further to 73.67\% for the complete test
set, confirms that satisfying chemical constraints becomes increasingly
difficult as molecular size increases. These findings further motivate the use
of postgeneration structural inspection, local graph repair, and
mass-spectral-evidence-based candidate reranking to improve both candidate
validity and identification accuracy.

\begin{table}[!t]
  \caption{Effect of LLM-assisted graph repair and reranking on the 1,490 test
  spectra containing no more than 10 heavy atoms. Parentheses give exact-match
  counts.}
  \label{tab:reranking}
  \centering
  \small
  \begin{tabular}{lcc}
    \toprule
    Metric & DiffGCMS & DiffGCMS + LLM \\
    \midrule
    Validity & 91.04\% & 100.00\% \\
    Acc@1 & 21.28\% (317) & 21.95\% (327) \\
    Acc@10 & 46.91\% (699) & 47.99\% (715) \\
    Tanimoto@1 & 39.94\% & 41.60\% \\
    Tanimoto@10 & 63.37\% & 64.79\% \\
    \bottomrule
  \end{tabular}
\end{table}

\begin{figure*}[!t]
  \centering
  \includegraphics[width=\textwidth,trim={20.40bp 64.80bp 9.60bp 90.00bp},clip]{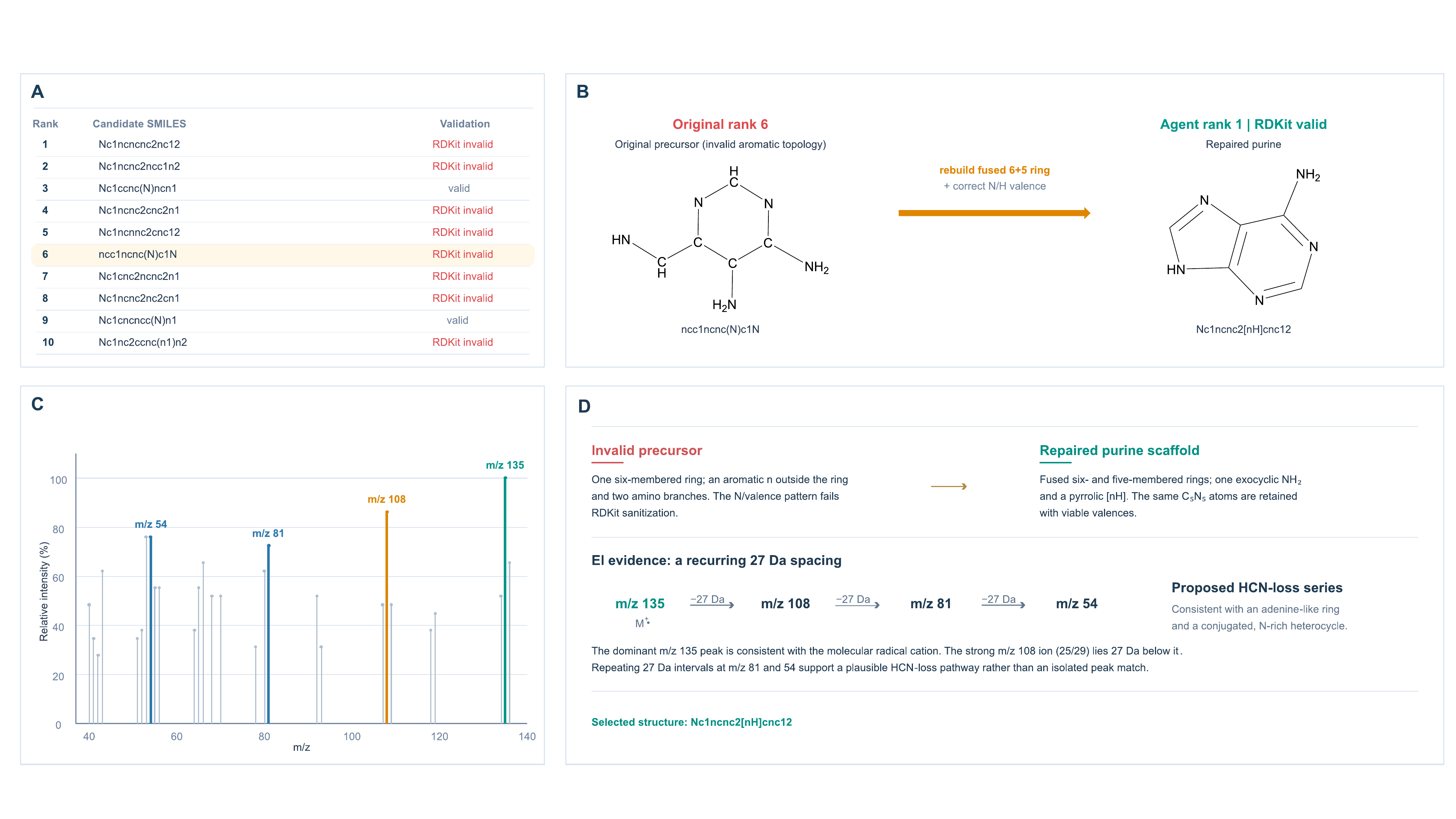}
  \caption{Representative example of molecular graph repair and spectrum-guided
  reranking of an aminopurine candidate. (A) Ten candidate SMILES generated by
  DiffGCMS and their RDKit validation outcomes; eight candidates were labeled
  invalid. (B) The candidate originally ranked sixth was repaired by
  reconstructing a fused six- and five-membered ring system and correcting
  nitrogen valence and hydrogen assignments while retaining the C$_5$N$_5$
  heavy-atom composition. The repaired structure, \texttt{Nc1ncnc2[nH]cnc12},
  passed RDKit validation and was ranked first by the agent. (C) The EI spectrum
  highlights ions at $m/z$ 135, 108, 81, and 54. (D) Structural comparison and a
  proposed HCN-loss series connect the repaired purine scaffold with the
  observed 27 Da spacings. The fragmentation assignments provide a plausible
  interpretation of the spectrum.}
  \label{fig:case}
\end{figure*}

\subsection{Effect of LLM Reranking}
On the evaluation subset comprising 1,490 spectra of molecules containing no
more than 10 heavy atoms, LLM-assisted graph repair and reranking improved all
evaluated metrics (Table~\ref{tab:reranking}). Following LLM-based structural
inspection, local graph repair, and RDKit validation, candidate validity
increased from 91.04\% to 100.00\%, an absolute gain of 8.96 percentage points.
This result indicates that the proposed workflow effectively eliminated
valence violations, bond-order errors, and other chemically invalid
representations produced by the diffusion model. Acc@1 increased from 21.28\%
to 21.95\%, with the number of reference structures correctly ranked first
increasing from 317 to 327. This corresponds to a gain of 0.67 percentage
points and 10 additional Top-1 exact matches. Acc@10 increased from 46.91\% to
47.99\%, and the number of spectra for which the reference structure appeared
among the 10 candidates increased from 699 to 715, corresponding to 16
additional reference molecules recovered and an absolute improvement of 1.08
percentage points.

Because each spectrum was associated with 10 candidate structures both before
and after LLM processing, with no validity-based filtering or replacement
sampling, the observed validity gain was not produced by discarding invalid
candidates. Reranking alone cannot change Acc@10. The improvement
in Acc@10 therefore primarily arises from molecular graph repair. Some original
candidates were already close to their corresponding reference structures in
terms of molecular scaffold or atomic composition but failed the strict
canonical-SMILES exact-matching criterion because of errors in molecular
connectivity, bond order, aromaticity, or valence assignment. Spectrum-guided
graph repair corrected these structural errors and restored such candidates to
structures identical to the corresponding reference molecules. In contrast,
the improvement in Acc@1 reflects the combined effects of graph repair and
spectrum-aware reranking. This result demonstrates that integrating evidence
from molecular-mass consistency, diagnostic fragment ions, plausible
bond-cleavage pathways, and neutral losses can promote candidates that are
better supported by the experimental spectrum to the top of the ranked list.

The fingerprint-similarity metrics exhibited the same trend. Tanimoto@1
increased from 39.94\% to 41.60\%, an improvement of 1.66 percentage points,
whereas Tanimoto@10 increased from 63.37\% to 64.79\%, an improvement of 1.42
percentage points. These gains indicate that the benefits of LLM processing
were not limited to a small number of newly obtained exact matches but extended
to the overall structural similarity between the candidate and reference
molecules. Even when the exact reference structure was not fully recovered,
the repaired and reranked candidates were generally closer to the reference
molecules in terms of molecular scaffolds or local functional groups.
Collectively, the concurrent improvements in chemical validity, exact-match
accuracy, and fingerprint similarity support the effectiveness of the
LLM-assisted stage: molecular graph repair eliminates invalid structures and
recovers otherwise missed exact matches, while spectrum-aware reranking
improves the priority assigned to the reference structure within the candidate
list.

\subsection{Interpretable Recovery Case}
Figure~\ref{fig:case} illustrates how LLM-assisted molecular graph repair and
spectrum-guided reranking can transform a chemically invalid candidate into a
valid aminopurine structure. Among the 10 candidate SMILES generated by
DiffGCMS, eight were labeled invalid by RDKit, whereas the candidates at ranks
3 and 9 were valid (Figure~\ref{fig:case}A). The candidate originally ranked
sixth, \texttt{ncc1ncnc(N)c1N}, was selected for the repair illustrated in
Figure~\ref{fig:case}B. As shown in the structural comparison, this candidate
contained a six-membered ring, an aromatic nitrogen outside the ring, and two
amino branches. Its invalid aromatic topology and nitrogen valence pattern
required correction before further structural evaluation.

The repair retained the C$_5$N$_5$ heavy-atom composition while revising the
molecular connectivity to construct a fused six- and five-membered purine ring
system (Figure~\ref{fig:case}B,D). Nitrogen valence and hydrogen assignments
were adjusted to accommodate one exocyclic amino group and a pyrrolic
\texttt{[nH]} within the fused heterocycle. The resulting SMILES,
\texttt{Nc1ncnc2[nH]cnc12}, passed RDKit validation. Thus, the transformation
involved reconstruction of the ring topology together with correction of the
nitrogen valence pattern, extending beyond a change in aromatic notation or
adjacent bond orders alone.

The EI spectrum provided complementary evidence for the repaired structure
(Figure~\ref{fig:case}C,D). The base peak at $m/z$ 135 was consistent with the
molecular radical cation of the C$_5$H$_5$N$_5$ structure. The prominent ion at
$m/z$ 108 lay 27 Da below this peak, and the ions at $m/z$ 81 and 54 extended
the same 27 Da spacing. These relationships were consistent with the proposed
HCN-loss series, $135 \rightarrow 108 \rightarrow 81 \rightarrow 54$, shown
in Figure~\ref{fig:case}D, providing a coherent fragmentation rationale for
the conjugated nitrogen-rich heterocycle. Because the observed mass spacings
alone do not establish a unique fragmentation sequence, this series is
presented as a plausible interpretation rather than a confirmed mechanism.

Following graph repair and spectrum-guided evaluation, the structure derived
from the original rank-6 candidate was assigned rank 1 by the agent
(Figure~\ref{fig:case}B). This example illustrates the complementary roles of
chemical validation and spectral interpretation: graph repair produced a valid
fused heterocycle, while the molecular-ion assignment and recurring 27 Da
spacings provided a traceable rationale for its prioritization. The case
therefore demonstrates how a chemically invalid generated candidate can be
converted into a valid structural hypothesis supported by an interpretable
account of the observed EI spectrum.

\FloatBarrier
\subsection{Scope, Limitations, and Future Perspectives}
Although DiffGCMS takes low-resolution, integer-valued $m/z$ peaks as input, its
candidate generation process relies heavily on a known molecular formula as a
structural constraint. Reliable determination of the molecular formula of an
unknown compound, however, generally requires the accurate-mass and isotopic-
distribution information provided by high-resolution mass spectrometry
\cite{kind2007,siriusformula,sirius4,schymanski2014}.
Therefore, the practical application of the current DiffGCMS generator
presupposes that a reliable molecular formula can be obtained using
high-resolution mass spectrometry or other experimental techniques. The model
is thus primarily applicable to analytical scenarios supported by
high-resolution mass spectrometry and cannot yet perform structure generation
for compounds with unknown molecular formulas solely from low-resolution EI
mass spectra.

Future studies could integrate the current mass-spectral evidence with
experimentally measured or model-predicted retention indices\cite{qu2021},
molecular-formula uncertainty, authentic-standard
validation, and other complementary information to improve the reliability of
structure identification. In addition, DeepSeek-V4-Flash was used as the
foundation model in the second stage of this study, with its reasoning mode
disabled. Enabling the reasoning mode or adopting a more capable foundation
model, such as GPT-6, may further improve the performance of the LLM in
candidate inspection, molecular graph repair, fragment-evidence interpretation,
and candidate reranking. Moreover, the second-stage LLM-assisted reasoning
workflow is not tied to the specific generative architecture of DiffGCMS. Its
general paradigm of candidate inspection, local graph repair, spectral-evidence
validation, and candidate reranking could, in principle, be transferred to
other spectroscopy-based molecular structure inference tasks, providing a
general framework for candidate selection and interpretable analysis across
different types of spectroscopic data.
\section{Conclusions}
This study presents a two-stage framework for de novo structure inference from
GC--EI--MS, combining a spectrum-conditioned discrete graph diffusion model with
LLM-assisted molecular graph repair and evidence-based reranking. DiffGCMS
generates formula-constrained molecular graphs by denoising interatomic
connectivity. In the second stage, candidate SMILES are converted into molecular
graphs and subjected to local repair while preserving their atomic composition.
RDKit is then used to validate their chemical validity, and the candidates are
reranked according to evidence from molecular ions, diagnostic fragments, bond
cleavages, and neutral losses. By separating candidate generation from chemical
and mass-spectral validation, the framework provides traceable peak-level
explanations for its predictions.

On the molecule-disjoint NIST 20 test set containing 13,696 spectra, DiffGCMS
achieved Acc@1 and Acc@10 values of 6.01\% and 15.76\%, respectively, together
with Tanimoto@1 and Tanimoto@10 values of 30.39\% and 45.77\%. On the subset of
1,490 spectra corresponding to molecules with no more than 10 heavy atoms,
LLM-assisted processing increased candidate validity from 91.04\% to 100.00\%,
Acc@1 from 21.28\% to 21.95\%, Acc@10 from 46.91\% to 47.99\%, Tanimoto@1 from
39.94\% to 41.60\%, and Tanimoto@10 from 63.37\% to 64.79\%. The case analysis
further illustrated how reconstruction of an invalid ring topology, while
preserving the heavy-atom composition, could yield a valid purine scaffold that
was prioritized through spectrum-guided reranking, with a proposed HCN-loss
series providing a traceable fragmentation rationale.

\section*{Associated Content}

\subsection*{Supporting Information}
Dataset construction and preprocessing; mathematical formulation of the diffusion model; candidate generation; Top-10 exact-match transitions following LLM-assisted processing; and additional spectrum-driven reranking and validation examples (PDF).

\section*{Author Information}
{\setlength{\parindent}{0pt}
\setlength{\parskip}{4pt}

\textbf{Corresponding Author}\par
\textbf{Fanyang Mo} --- \textit{School of AI for Science, Peking University
Shenzhen Graduate School, Shenzhen 518055, China; State Key Laboratory of
Advanced Waterproof Materials, School of Materials Science and Engineering,
Peking University, Beijing 100871, China; School of Advanced Materials, Peking
University Shenzhen Graduate School, Shenzhen 518055, China; Guangdong
Provincial Key Laboratory of Nano-Micro Materials Research, Peking University
Shenzhen Graduate School, Shenzhen 518055, China}; Email:
\href{mailto:fmo@pku.edu.cn}{fmo@pku.edu.cn}

\textbf{Authors}\par
\textbf{Changlin Liu} --- \textit{School of AI for Science, Peking University
Shenzhen Graduate School, Shenzhen 518055, China; School of Advanced Materials,
Peking University Shenzhen Graduate School, Shenzhen, 518055, China}; ORCID:
\href{https://orcid.org/0009-0007-1234-4192}{0009-0007-1234-4192}; Email:
\href{mailto:changlinliu25@stu.pku.edu.cn}{changlinliu25@stu.pku.edu.cn}

\textbf{Tianyu Yi} --- \textit{School of AI for Science, Peking University
Shenzhen Graduate School, Shenzhen 518055, China; School of Advanced Materials,
Peking University Shenzhen Graduate School, Shenzhen, 518055, China}

\textbf{Chengchun Liu} --- \textit{School of AI for Science, Peking University
Shenzhen Graduate School, Shenzhen 518055, China; State Key Laboratory of
Advanced Waterproof Materials, School of Materials Science and Engineering,
Peking University, Beijing 100871, China}

\textbf{Boxuan Zhao} --- \textit{School of AI for Science, Peking University
Shenzhen Graduate School, Shenzhen 518055, China; State Key Laboratory of
Advanced Waterproof Materials, School of Materials Science and Engineering,
Peking University, Beijing 100871, China}

\textbf{Author Contributions}\par
The manuscript was written through contributions of all authors. All authors
have given approval to the final version of the manuscript.

\textbf{Notes}\par
The authors declare no competing financial interest.
}

\section*{Acknowledgments}
This work was supported by the New Generation Artificial Intelligence-National Science and Technology Major Project (Grant No. 2026ZD0128504), as well as by the National Natural Science Foundation of China through the Young Scientists Fund-Type A (the former National Science Fund for Distinguished Young Scholars, NSFDYS) (Grant No. T262500040) and the General Program (Grant No. 22673006). We gratefully acknowledge Peking University Shenzhen Graduate School and the Shenzhen Government for start-up funding support. We also thank the High-Performance Computing Platform of Peking University for providing computational resources for machine-learning model training.

\clearpage
\onecolumn
\section*{Table of Contents Graphic}
\noindent\includegraphics[width=3.25in,height=1.75in,keepaspectratio]{fig1.pdf}


{\footnotesize
\begin{thebibliography}{99}
\setlength{\itemsep}{0pt}
\setlength{\parskip}{0pt}

\bibitem{nist20}
National Institute of Standards and Technology. \textit{NIST/EPA/NIH Mass
Spectral Library with Search Program, NIST Standard Reference Database 1A,
Version 20}; NIST: Gaithersburg, MD, 2020. DOI: 10.18434/T4H594.

\bibitem{stein2012}
Stein, S. E. Mass Spectral Reference Libraries: An Ever-Expanding Resource for
Chemical Identification. \textit{Anal. Chem.} \textbf{2012}, \textit{84},
7274--7282. DOI: 10.1021/ac301205z.

\bibitem{stein1994}
Stein, S. E.; Scott, D. R. Optimization and Testing of Mass Spectral Library
Search Algorithms for Compound Identification. \textit{J. Am. Soc. Mass
Spectrom.} \textbf{1994}, \textit{5}, 859--866. DOI:
10.1016/1044-0305(94)87009-8.

\bibitem{stein1999}
Stein, S. E. An Integrated Method for Spectrum Extraction and Compound
Identification from Gas Chromatography/Mass Spectrometry Data. \textit{J. Am.
Soc. Mass Spectrom.} \textbf{1999}, \textit{10}, 770--781. DOI:
10.1016/S1044-0305(99)00047-1.

\bibitem{massbank}
Horai, H.; et al. MassBank: A Public Repository for Sharing Mass Spectral Data
for Life Sciences. \textit{J. Mass Spectrom.} \textbf{2010}, \textit{45},
703--714. DOI: 10.1002/jms.1777.

\bibitem{metfrag}
Wolf, S.; Schmidt, S.; M\"{u}ller-Hannemann, M.; Neumann, S. In Silico
Fragmentation for Computer Assisted Identification of Metabolite Mass Spectra.
\textit{BMC Bioinf.} \textbf{2010}, \textit{11}, 148. DOI:
10.1186/1471-2105-11-148.

\bibitem{cfmei}
Allen, F.; Pon, A.; Greiner, R.; Wishart, D. Computational Prediction of Electron
Ionization Mass Spectra to Assist in GC/MS Compound Identification.
\textit{Anal. Chem.} \textbf{2016}, \textit{88}, 7689--7697. DOI:
10.1021/acs.analchem.6b01622.

\bibitem{krettler2021}
Krettler, C. A.; Thallinger, G. G. A Map of Mass Spectrometry-Based In Silico
Fragmentation Prediction and Compound Identification in Metabolomics.
\textit{Brief. Bioinform.} \textbf{2021}, \textit{22}, bbab073. DOI:
10.1093/bib/bbab073.

\bibitem{spec2mol}
Litsa, E. E.; Chenthamarakshan, V.; Das, P.; Kavraki, L. E. An End-to-End Deep
Learning Framework for Translating Mass Spectra to De-Novo Molecules.
\textit{Commun. Chem.} \textbf{2023}, \textit{6}, 132. DOI:
10.1038/s42004-023-00932-3.

\bibitem{massgenie}
Shrivastava, A. D.; Swainston, N.; Samanta, S.; Roberts, I.; Wright Muelas, M.;
Kell, D. B. MassGenie: A Transformer-Based Deep Learning Method for Identifying
Small Molecules from Their Mass Spectra. \textit{Biomolecules} \textbf{2021},
\textit{11}, 1793. DOI: 10.3390/biom11121793.

\bibitem{msnovelist}
Stravs, M. A.; D\"{u}hrkop, K.; B\"{o}cker, S.; Zamboni, N. MSNovelist: De Novo
Structure Generation from Mass Spectra. \textit{Nat. Methods} \textbf{2022},
\textit{19}, 865--870. DOI: 10.1038/s41592-022-01486-3.

\bibitem{ms2mol}
Butler, T.; Frandsen, A.; Lightheart, R.; Bargh, B.; Kerby, T.; West, K.;
et al. MS2Mol: A Transformer Model for Illuminating Dark Chemical Space
from Mass Spectra. \textit{ChemRxiv} \textbf{2023}. DOI:
10.26434/chemrxiv-2023-vsmpx-v4.

\bibitem{mass2smiles}
Elser, D.; Huber, F.; Gaquerel, E. Mass2SMILES: Deep Learning Based Fast
Prediction of Structures and Functional Groups Directly from High-Resolution
MS/MS Spectra. \textit{bioRxiv} \textbf{2023}. DOI:
10.1101/2023.07.06.547963.

\bibitem{cyclems}
Lin, J.; Li, H.; Liu, C.; Yan, Z.; Yuan, L.; Mo, F.
Cycle-MS: A Closed-Loop End-to-End Framework for Mass Spectrometry
Structure Elucidation. \textit{J. Chem. Inf. Model.}
\textbf{2026}, \textit{66}, 2709--2718.
DOI: 10.1021/acs.jcim.5c02981.

\bibitem{diffms}
Bohde, M.; Manjrekar, M.; Wang, R.; Ji, S.; Coley, C. W. DiffMS: Diffusion
Generation of Molecules Conditioned on Mass Spectra. In \textit{Proceedings of
the 42nd International Conference on Machine Learning}; Proceedings of Machine
Learning Research, 2025; Vol. 267, pp 4737--4756.
\url{https://proceedings.mlr.press/v267/bohde25a.html}.

\bibitem{flowms}
Nie, J.; Gao, P. FlowMS: Flow Matching for De Novo Structure Elucidation from
Mass Spectra. \textit{arXiv} \textbf{2026}, arXiv:2603.18397. DOI:
10.48550/arXiv.2603.18397.

\bibitem{neims}
Wei, J. N.; Belanger, D.; Adams, R. P.; Sculley, D. Rapid Prediction of
Electron--Ionization Mass Spectrometry Using Neural Networks. \textit{ACS Cent.
Sci.} \textbf{2019}, \textit{5}, 700--708. DOI:
10.1021/acscentsci.9b00085.

\bibitem{rassp}
Zhu, R. L.; Jonas, E. Rapid Approximate Subset-Based Spectra Prediction for
Electron Ionization--Mass Spectrometry. \textit{Anal. Chem.} \textbf{2023},
\textit{95}, 2653--2663. DOI: 10.1021/acs.analchem.2c02093.

\bibitem{deepei}
Ji, H.; Deng, H.; Lu, H.; Zhang, Z. Predicting a Molecular Fingerprint from an
Electron Ionization Mass Spectrum with Deep Neural Networks. \textit{Anal.
Chem.} \textbf{2020}, \textit{92}, 8649--8653. DOI:
10.1021/acs.analchem.0c01450.

\bibitem{spectus}
H\'{a}jek, A.; Star\'{y}, M.; Price, E.; Jozefov, F.; Hecht, H.;
K\v{r}enek, A. SpecTUS: Spectral Translator for Unknown Structures Annotation
from EI-MS Spectra. \textit{Anal. Chem.} \textbf{2026}, \textit{98},
21670--21682. DOI: 10.1021/acs.analchem.6c02423.

\bibitem{rdkit}
Landrum, G.; RDKit Contributors. \textit{RDKit: Open-Source Cheminformatics
Software}, version 2022.09.5 (Release\_2022\_09\_5); Zenodo, 2023.
DOI: 10.5281/zenodo.7671152.

\bibitem{smiles}
Weininger, D. SMILES, a Chemical Language and Information System. 1.
Introduction to Methodology and Encoding Rules. \textit{J. Chem. Inf. Comput.
Sci.} \textbf{1988}, \textit{28}, 31--36. DOI: 10.1021/ci00057a005.

\bibitem{d3pm}
Austin, J.; Johnson, D. D.; Ho, J.; Tarlow, D.; van den Berg, R. Structured
Denoising Diffusion Models in Discrete State-Spaces. In \textit{Advances in
Neural Information Processing Systems}; 2021; Vol. 34, pp 17981--17993. DOI:
10.48550/arXiv.2107.03006.

\bibitem{digress}
Vignac, C.; Krawczuk, I.; Siraudin, A.; Wang, B.; Cevher, V.; Frossard, P.
DiGress: Discrete Denoising Diffusion for Graph Generation. In
\textit{International Conference on Learning Representations}; 2023. DOI:
10.48550/arXiv.2209.14734.

\bibitem{nichol2021}
Nichol, A. Q.; Dhariwal, P. Improved Denoising Diffusion Probabilistic Models.
In \textit{Proceedings of the 38th International Conference on Machine
Learning}; 2021; pp 8162--8171. DOI: 10.48550/arXiv.2102.09672.

\bibitem{graphtransformer}
Dwivedi, V. P.; Bresson, X. A Generalization of Transformer Networks to Graphs.
\textit{arXiv} \textbf{2020}, arXiv:2012.09699. DOI:
10.48550/arXiv.2012.09699.

\bibitem{adamw}
Loshchilov, I.; Hutter, F. Decoupled Weight Decay Regularization. In
\textit{International Conference on Learning Representations}; 2019. DOI:
10.48550/arXiv.1711.05101.

\bibitem{onecycle}
Smith, L. N.; Topin, N. Super-Convergence: Very Fast Training of Neural
Networks Using Large Learning Rates. \textit{arXiv} \textbf{2017},
arXiv:1708.07120. DOI: 10.48550/arXiv.1708.07120.

\bibitem{molecode}
Yan, Z.; Liu, C.; Zhao, B.; Lin, K.; Zhao, J.; Wang, Y.; Lv, L.; Li, H.;
Zhang, S.; Yuan, L.; Mo, F. MoleCode Unlocks Structural Intelligence in Large
Language Models. \textit{arXiv} \textbf{2026}, arXiv:2605.16480. DOI:
10.48550/arXiv.2605.16480.

\bibitem{chemcrow}
Bran, A. M.; Cox, S.; Schilter, O.; Baldassari, C.; White, A. D.; Schwaller, P.
Augmenting Large Language Models with Chemistry Tools.
\textit{Nat. Mach. Intell.} \textbf{2024}, \textit{6}, 525--535.
DOI: 10.1038/s42256-024-00832-8.

\bibitem{ecfp}
Rogers, D.; Hahn, M. Extended-Connectivity Fingerprints. \textit{J. Chem. Inf.
Model.} \textbf{2010}, \textit{50}, 742--754. DOI: 10.1021/ci100050t.

\bibitem{kind2007}
Kind, T.; Fiehn, O. Seven Golden Rules for Heuristic Filtering of Molecular
Formulas Obtained by Accurate Mass Spectrometry. \textit{BMC Bioinf.}
\textbf{2007}, \textit{8}, 105. DOI: 10.1186/1471-2105-8-105.

\bibitem{siriusformula}
B\"{o}cker, S.; Letzel, M. C.; Lipt\'{a}k, Z.; Pervukhin, A. SIRIUS:
Decomposing Isotope Patterns for Metabolite Identification.
\textit{Bioinformatics} \textbf{2009}, \textit{25}, 218--224. DOI:
10.1093/bioinformatics/btn603.

\bibitem{sirius4}
D\"{u}hrkop, K.; Fleischauer, M.; Ludwig, M.; Aksenov, A. A.; Melnik, A. V.;
Meusel, M.; Dorrestein, P. C.; Rousu, J.; B\"{o}cker, S. SIRIUS 4: A Rapid Tool
for Turning Tandem Mass Spectra into Metabolite Structure Information.
\textit{Nat. Methods} \textbf{2019}, \textit{16}, 299--302. DOI:
10.1038/s41592-019-0344-8.

\bibitem{schymanski2014}
Schymanski, E. L.; Jeon, J.; Gulde, R.; Fenner, K.; Ruff, M.; Singer, H. P.;
Hollender, J. Identifying Small Molecules via High Resolution Mass
Spectrometry: Communicating Confidence. \textit{Environ. Sci. Technol.}
\textbf{2014}, \textit{48}, 2097--2098. DOI: 10.1021/es5002105.

\bibitem{qu2021}
Qu, C.; Schneider, B. I.; Kearsley, A. J.; Keyrouz, W.; Allison, T. C.
Predicting Kov\'{a}ts Retention Indices Using Graph Neural Networks.
\textit{J. Chromatogr. A} \textbf{2021}, \textit{1646}, 462100.
DOI: 10.1016/j.chroma.2021.462100.

\end{thebibliography}
}
\end{document}